\documentclass[12pt,amsmath,amssymb]{revtex4}
\usepackage{graphics}
\usepackage{amsfonts}
\usepackage{amsmath}
\usepackage{epsfig}
\usepackage{epsf}
\usepackage{graphicx}
\usepackage{dcolumn}
\usepackage{bm}
\usepackage{color}
\usepackage{array}
\usepackage{amssymb}

\usepackage{breqn}

\newcommand {\sla}[1]{ #1 \!\!\!/}

\begin{document}
\title{Correction to the energy spectrum of $^1S_0$ heavy quarkonia due to two-gluon annihilation effect}
\author{
Hui-Yun Cao, Hai-Qing Zhou \protect\footnotemark[1] \protect\footnotetext[1]{E-mail: zhouhq@seu.edu.cn} \\
School of Physics,
Southeast University, NanJing 211189, China}
\date{\today}

\begin{abstract}
In this work, the non-relativistic asymptotic behavior of the transition $q\overline{q}\rightarrow2g\rightarrow q\overline{q}$ in the $^1S_0$ channel is discussed. Different with the usual calculation which expands the physical amplitude around the quark anti-quark threshold, we take the quark anti-quark pairs as off shell and only expand the expression on the three-dimensional momenta of the quarks and anti-quarks. We calculate the results to order 6. The imagine part of the results after applying the on shell conditions can reproduce the non-relativistic QCD (NRQCD) results in leading order of $\alpha_s$. The real part of the results can be used to estimate the mass shift of the $^1S_0$ heavy quark anti-quark system due to the $2g$ annihilation effect. The results can also be used to estimate the energy shifts of the positronium system due to the two-photon annihilation.
\end{abstract}

\maketitle

\section{\label{sec-1}Introduction}

The energy spectrum of quarkonia is a basic topic of the strong interaction. Many phenomenological methods have been applied to study this topic for a long time such as quark model\cite{quark-model}, QCD sum rules\cite{QCD-sumrule}, Dyson-Schwinger equation and Bethe-Salpeter equation \cite{BS-eq} and unitary chiral model\cite{chiral-unitary} \emph{etc.}. Due to the asymptotic freedom of the QCD and the large masses of the heavy quarks ($c$ quark and $b$ quark), the heavy quarkonia provides a special window to study the QCD since both the pertuabtive behavior and the non-perturbative behavior show their properties in such systems. For example, the spectrum of heavy quarkonia shows the non-perturbative confinement behavior, on another hand the decay and the production of the heavy quarkonia can be well described by the effective theory non-relativistic QCD (NRQCD)\cite{NRQCD}.

Experimentally, since 2003 the Belle\cite{Belle}, CDF\cite{CDF}, D0\cite{D0}, BarBar\cite{Barbar}, Cleo-C\cite{Cleo-C}, LHCb\cite{LHCb}, BES\cite{BESIII} and CMS\cite{CMS} collaborations have reported many new charmonium-like states which can not be understood well even in the phenomenological level. It is found for the states below the threshold of $D$ or $D^*$ pair the experimental results and theoretical calculations are compatible, while above the threshold of $D$ or $D^*$ pair the situation is perplexing. This attracts a lot of interest from both theoretical and experimental physicists and numerous studies are tried to understand these states. The detail of these discussion can be found in the recent reviews \cite{recent-review}. Physically, when the masses of the states lie above the threshold of $D$ or $D^*$ pair, the corresponding decay channels are opened. The interactions related with these decay channels not only result in the decay widths, but also shift the masses. A natural and important question is how large are these effects and how to estimate them. The effects to the energy spectrum of charnonium due to the decay channels (annihilation effects) have been discussed in the original quark model \cite{quark-model} and subsequent work in a phenomenological way \cite{Barnes2008}. The contributions are expected to be about $20$ Mev in the former and are about $600$ Mev in the later. In this work, we plan to give a rigorous study on the similare effects due to the two-gluon annihilation which is much clear than the $D\overline{D},D\overline{D}^*, D^*\overline{D}^*$ decay channels and can be well described in a pure perturbative frame.

In the quasi potential method, the effective potential of a non-relativistic system is usually extracted from the matching between the quantum mechanics and the full quantum field theory by expanding the physical amplitudes on the threshold order by order. The similar matching conditions are usually also applied between the effective theory and the full theory such as NRQCD and QCD which means the two theories are equivalent in the physical scattering region. To study the corrections to the energy shifts of bound states due to the two-gluon annihilation, in this work we do not match the on shell amplitudes but take the quark anti-quark pairs as off shell and then expand the interaction kernel order by order. The gauge invariance of such expansion is also check in a manifest way.

We organize the paper as following, in section II we give an introduction on the basic formula, in section III we describe the way of our calculation and present the analytic result for the coefficients to order 6 after the non-relativistic expansion, in section IV we discuss the relation between our results and those given in NRQCD in the leading order of $\alpha_s$ (LO-$\alpha_s$), in section V we estimate the effects to the mass shifts numerically and give our conclusion.

\section{\label{sec-2} Basic formula}

\begin{figure}[htbp]
\center{\epsfxsize 3.3 truein\epsfbox{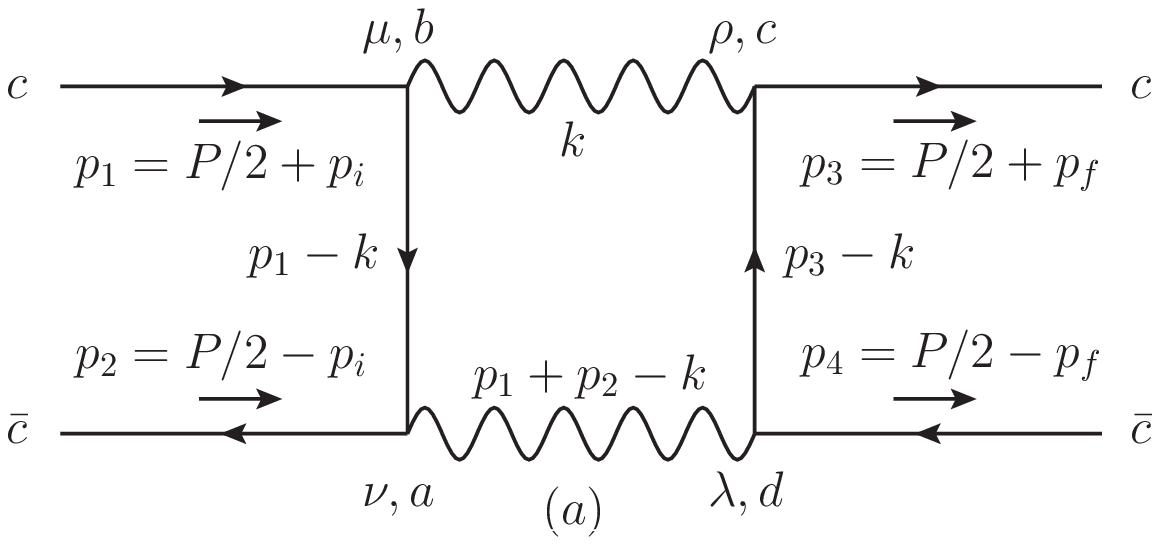}\epsfxsize 3.3 truein\epsfbox{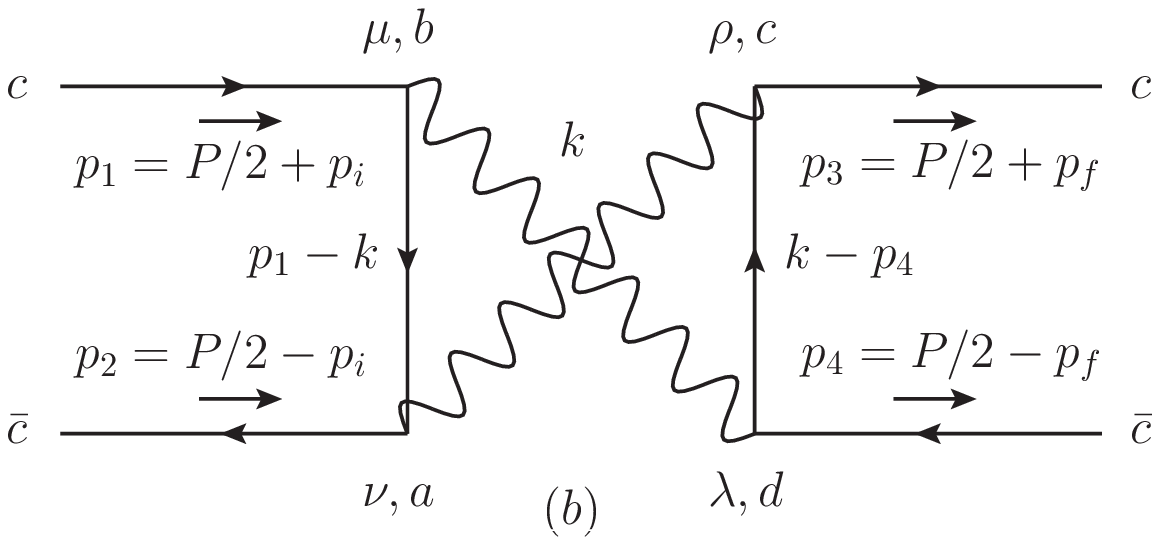}}
\caption{The Feynman diagrams for $q\overline{q}\rightarrow 2g\rightarrow q\overline{q}$.}
\label{figure:TGE}
\end{figure}

In the perturbation theory, the Feynman diagrams for the transition of a heavy quark anti-quark pair to a heave quark anti-quark pair via two-gluon annihilation $q\overline{q}\rightarrow 2g\rightarrow q\overline{q}$ are showed as Fig. \ref{figure:TGE}.

\begin{figure}[htbp]
\center{\epsfxsize 3.3 truein\epsfbox{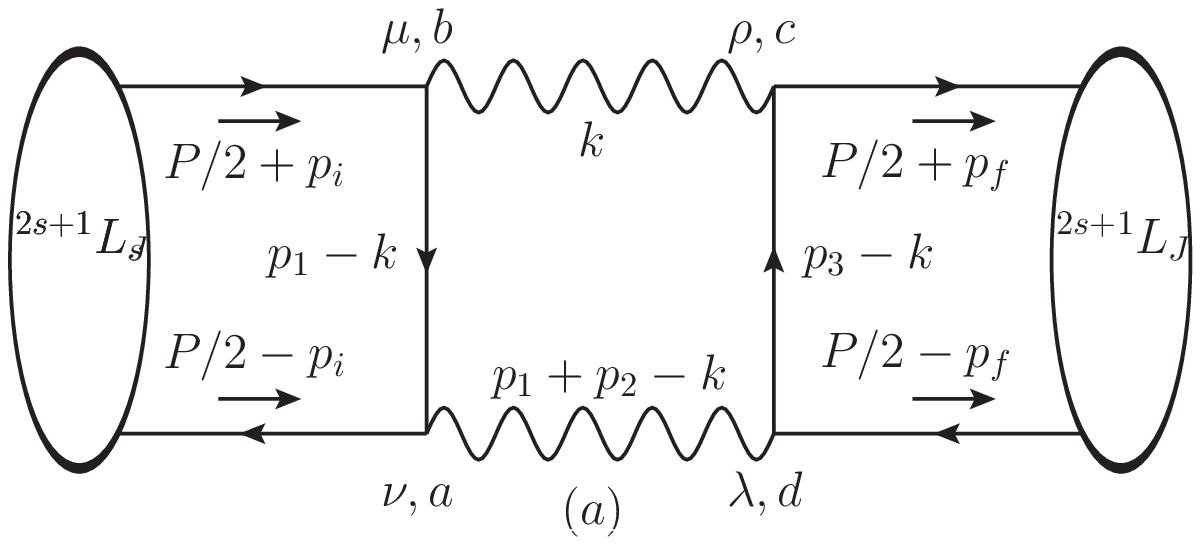}\epsfxsize 3.3 truein\epsfbox{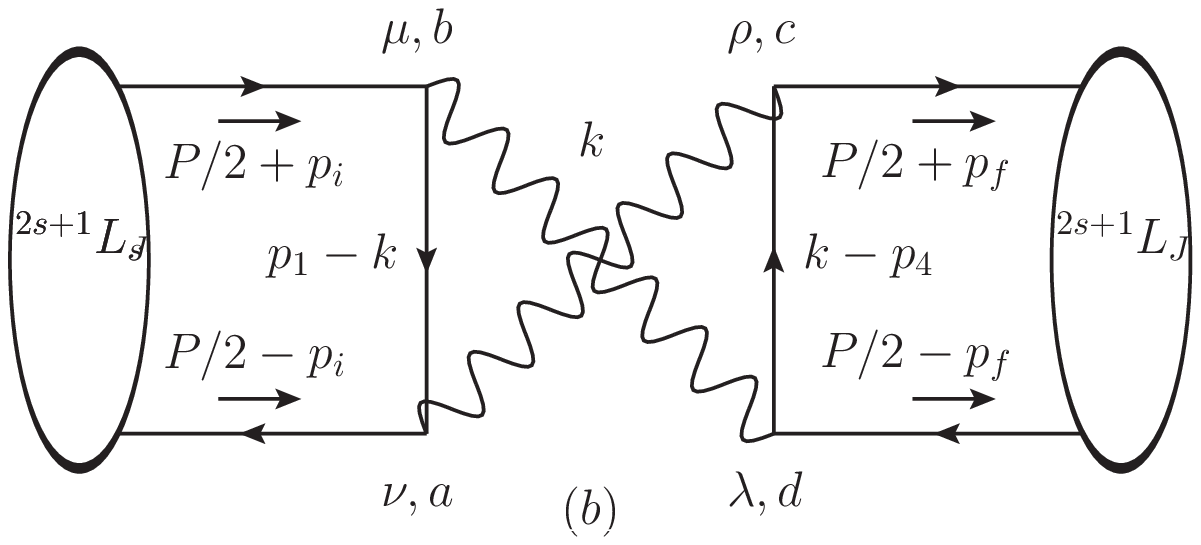}}
\caption{The Feynman diagrams for $q\overline{q}\rightarrow 2g\rightarrow q\overline{q}$ in $^{2s+1}L_J$ channel.}
\label{figure:TGE-project}
\end{figure}

When one take the quark anti-quark pairs as off shell, the corresponding Green function is a part of the interaction kernel of the Bethe-Salpeter equation which plays the role like the potential in the non-relativistic quantum mechanics. The direct calculation of the corresponding Green function is a little tedious and there is no analytic expression in the full complex plane of momenta. Two methods are usually used to simplify the calculation. The first one is to project the quark anti-quark pairs to a special $^{2s+1}L_J$ state which is described as Fig. \ref{figure:TGE-project} and the second one is to study the asymptotic behavior of this Green function which means to expand the expression on some small variables.

In the center mass frame, the momenta can be chosen as following.
\begin{eqnarray}
p_1 &\triangleq& \frac{1}{2}P+p_i,~~~~p_2 \triangleq \frac{1}{2}P-p_i, \nonumber \\
p_3 &\triangleq& \frac{1}{2}P+p_f,~~~~p_4 \triangleq \frac{1}{2}P-p_f,
\end{eqnarray}
with $P \triangleq (\sqrt{s},0,0,0)$. In the general case, there are six independent Lorentz invariant variables in the interaction kernel: $P^2,~P\cdot p_i,~P\cdot p_f,~p_i^2,~p_f^2$. For simplicity, we limit our discussion in the case with $P\cdot p_i=P\cdot p_f=0$. This region is corresponding to the instantaneous approximation for the Bethe-Salpeter equation which means the contributions from the relative energy of the quark anti-quark pair in the bound states is neglected. This property naturally appears when the initial and final quark anti-quark pairs are taken as on shell. Such choice of the momenta leaves the number of the independent  Lorentz invariant variables to 4 and we can define $p_i\triangleq(0,\textbf{p}_i),p_f\triangleq(0,\textbf{p}_f)$.  This is different with the on shell case where there are only two independent Lorentz invariant variables $P^2 = s \triangleq 4m^2+4p^2$ and $p_i\cdot p_f$ with $m$ the mass of quark. For the heavy quark anti-quark pair system, we can take $p_i^2/m^2,~p_f^2/m^2,~p_i\cdot p_f/m^2 $ as small variables and leave $s$ as a free variable at first.

To project the quark anti-quark pairs to a special $^1S_0$ state, we use the same project matrix as the on shell case \cite{project-operator-1,project-operator-2,project-operator-3} where one has
\begin{eqnarray}
\sum \overline{v}(p_2,s_2) T u(p_1,s_1)<\frac{1}{2}s_1;\frac{1}{2}s_2|00> & \triangleq& \textrm{Tr}[T.\Pi_{ini}], \nonumber\\
\sum \overline{u}(p_3,s_3) T v(p_4,s_4)<\frac{1}{2}s_3;\frac{1}{2}s_4|00> & \triangleq & \textrm{Tr}[T.\Pi_{fin}],
\end{eqnarray}
where the Clebsch-Gordan coefficients are the standard ones as Ref. \cite{project-operator-2} and the Dirac spinors are normalized as $u^+u=v^+v=1$ whose expressions are written as
\begin{eqnarray}
u(p_1,s_1) &\triangleq&  \frac{\sla{p_}_1+m}{ \sqrt{E_1(E_1+ m)}} \begin{pmatrix} \xi^{s_1}\\0 \end{pmatrix}, \nonumber \\
v(p_2,s_2)&\triangleq& \frac{-\sla{p_}_2+m}{  \sqrt{E_2(E_2+ m)}} \begin{pmatrix} 0\\ \eta^{s_2} \end{pmatrix},
\end{eqnarray}
with $E_{1,2}\triangleq\sqrt{\textbf{p}_{1,2}^2+m^2}$, $\xi^{1/2}= (1,0)^T$, $\xi^{-1/2}= (0,1)^T$, $\eta^{1/2}= (0,1)^T$ and $\eta^{-1/2}= (-1,0)^T$. This results in the following expressions.
\begin{eqnarray}
\Pi_{ini} &=&  \frac{1}{8\sqrt{2} E_i^2 (E_i + m)}
(\sla{p}_1+m)(\sqrt{s}+\sla{P})\gamma^{5}(-\sla{p}_2+m),\nonumber \\
\Pi_{fin} &=& -\frac{1}{8\sqrt{2} E_f^2 (E_f + m)} (-\sla{p}_4+m)\gamma^{5}(\sqrt{s}+\sla{P})(\sla{p}_3+m),
\end{eqnarray}
where $E_{i,f}\triangleq \sqrt{\textbf{p}_{i,f}^2+m^2}$ and one should note that there is a minus in the expression of $\Pi_{fin}$.

We want to point out, the form of such project matrix is just for simplicity in our calculation. In principle the project matrix should be deduced from the Bethe-Salpeter equation and in the ultra non-relativistical limit the above expressions are expected to be true. In this work, we do not go to discuss the detail of this project matrix but just take the same form as the references.

Using the above project method, the interaction kernel in $^1S_0$ state can be expressed as the following.
\begin{eqnarray}
G^{(a)}(^1S_0)&=& -ic_f^{(2g)}\int \frac{d^D k}{(2\pi)^4} \textrm{Tr}[T_1 \cdot \Pi_{ini}] \textrm{Tr}[T_2 \cdot \Pi_{fin}]  D_{\mu\rho}(k)D_{\nu\lambda}(p_1+p_2-k),  \nonumber\\
G^{(b)}(^1S_0)&=& -ic_f^{(2g)} \int  \frac{d^D k}{(2\pi)^4} \textrm{Tr}[T_1 \cdot \Pi_{ini}] \textrm{Tr}[T_3 \cdot \Pi_{fin}]  D_{\mu\lambda} (k)D_{\nu\rho} (p_1+p_2-k),
\label{Gab}
\end{eqnarray}
where the color factor $c_f^{(2g)}$ is
\begin{eqnarray}
c_f^{(2g)}&=&(\frac{\delta_{ij}}{\sqrt{N_c}} T_a^{jm} T_b^{mi}) (\frac{\delta_{i'j'}}{\sqrt{N_c}} T_c^{j'm'} T_d^{m'i'}) \delta_{ad}\delta_{bc} \nonumber \\
&=&\frac{C_A C_F}{2N_c}=\frac{N_c^2-1}{4N_c}=\frac{2}{3},
\end{eqnarray}
and the hard kernel $T_i$ are
\begin{eqnarray}
T_1 &=& (-ig_s \gamma^{\nu}) \cdot S_F(p_1-k) \cdot  (-ig_s \gamma^\mu), \nonumber\\
T_2 &=& (-ig_s \gamma^\rho) \cdot  S_F(p_3-k)  \cdot (-ig_s \gamma^{\lambda}), \nonumber\\
T_3 &=& (-ig_s \gamma^{\rho}) \cdot S_F(k-p_4) \cdot (-ig_s \gamma^{\lambda}),
\end{eqnarray}
with
\begin{eqnarray}
S_F(q) &=& \frac{i(\sla{q}+m)}{q^2-{m}^2+i\epsilon}, \nonumber \\
D_{\mu\rho}(q) &=& \frac{-i }{q^2+i\epsilon}(g_{\mu\rho}- \xi \frac{q_\mu q_\rho}{q^2}).
\end{eqnarray}

For the on shell quark anti-quark pairs one has $(P/2\pm p_i)^2=(P/2 \pm p_f)^2=m^2$ and these conditions lead to the denominators of the integrands in Eq. (\ref{Gab}) include terms like $k^2-(P-2p_{i,f})\cdot k +i\epsilon$. This situation leads to the expansion on  $p_{i,f}$ and the integration of the loop momentum un-commutative when one goes to estimate the asymptotic behavior of the expression. In this case, one can separate the loop integration into hard part and other parts using the region method \cite{region-methos}. For the off shell quark anti-quark pairs, $(P^2/4-m^2)$ is taken as a free finite quantity and one can commutate the expansion and the integration of the loop momentum safely.

\section{\label{sec-3} The analytic results for the asymptotic behavior}
In our calculation, we at first use the package Feyncalc \cite{FeynCalc} to do the trace of Dirac matrixes in $D$-dimension as Ref. \cite{JiaYu2011}, then expand the expression on the variables $p_i^2,~p_f^2,~p_i\cdot p_f$ to a special order. This expansion is equivalent to expand the expression on the four momenta $p_i$ and $p_f$ directly. After the expansion, we use the tensor decomposition  to re-expressed the loop integrations as following.
\begin{eqnarray}
\int_k  \frac{(p_{i}\cdot k)^6}{k^2(P-k)^2((P/2-k)^2-m^2)^n}&=& \int_k \frac{ -15[k^2s-(k \cdot P)^2]^3 |\textbf{p}_i|^6}{(D+3)(D^2-1)s^3k^2(P-k)^2((P/2-k)^2-m^2)^n}, \nonumber \\
\int_k  \frac{(p_{i}\cdot k)^4}{k^2(P-k)^2((P/2-k)^2-m^2)^n} &=& \int_k \frac{3[(k \cdot P)^2-k^2s]^2|\textbf{p}_i|^4}{(D^2-1)s^2k^2(P-k)^2((P/2-k)^2-m^2)^n}, \nonumber \\
\int_k \frac{(p_{i}\cdot k)^2}{k^2(P-k)^2((P/2-k)^2-m^2)^n} &=& \int_k \frac{[sk^2-(k\cdot P)^2] |\textbf{p}_i|^2}{(1-D)sk^2(P-k)^2((P/2-k)^2-m^2)^n}, \nonumber \\
\int_k  \frac{(p_{i}\cdot k)^4 \ (p_{f}\cdot k)^2}{k^2(P-k)^2((P/2-k)^2-m^2)^n}&=& \int_k \frac{-3 (4\beta^2+1) [k^2s-(k \cdot P)^2]^3|\textbf{p}_i|^4|\textbf{p}_f|^2}{(D+3)(D^2-1)s^3k^2(P-k)^2((P/2-k)^2-m^2)^n}. \nonumber \\
\int_k  \frac{(p_{i}\cdot k)^2 \ (p_{f}\cdot k)^2}{k^2(P-k)^2((P/2-k)^2-m^2)^n}&=& \int_k \frac{ (2\beta^2+1) [(k \cdot P)^2-k^2s]^2|\textbf{p}_i|^2|\textbf{p}_f|^2}{(D^2-1)s^2k^2(P-k)^2((P/2-k)^2-m^2)^n},
\end{eqnarray}
with $\beta \triangleq \textbf{p}_i.\textbf{p}_f/ |\textbf{p}_i| |\textbf{p}_f|$ and $\int_k \triangleq \int d^Dk$. After the expansion and the tensor decomposition, for simplicity we directly use the package FIESTA \cite{FIESTA} to do the sector decomposition with $i\epsilon$ kept in the propagators and output the date base for integration, then use Mathematica to do the analytic integration.

In the practical calculation, we expand the expressions on $|\textbf{p}_{i,f}|$ to order 6. The gauge parameter $\xi$ is also kept in the practical calculation and we find the result is not dependent on $\xi$ which means the result is gauge independent although the momentum $P/2\pm p_{i,f}$ are not on shell. The direct numerical calculation is also used to check the analytic result.

After the loop integration, we apply the following property to reduce the variables since we only care for the matrix element of the interaction kernel between the $^1S_0$ states.
\begin{eqnarray}
\int d\Omega_{\textbf{p}_i}d\Omega_{\textbf{p}_f} \textbf{p}_i\cdot \textbf{p}_f  &=& 0,\nonumber \\
\int d\Omega_{\textbf{p}_i}d\Omega_{\textbf{p}_f} (\textbf{p}_i\cdot \textbf{p}_f)^2 &=& \frac{1}{3} |\textbf{p}_i| |\textbf{p}_f| \int d\Omega_{\textbf{p}_i}d\Omega_{\textbf{p}_f}.
\label{eq:pipf}
\end{eqnarray}
The high terms like $(\textbf{p}_i\cdot \textbf{p}_f)^4$ are not appeared in the expression and we need not care for them.

Eq.(\ref{eq:pipf}) means we can replace the terms $(\textbf{p}_i\cdot \textbf{p}_f)^2$ and $(\textbf{p}_i\cdot \textbf{p}_f)$ by $\frac{1}{3} |\textbf{p}_i| |\textbf{p}_f|$ and $0$ in our discussion. After such replacement, the final result can be expressed as
\begin{eqnarray}
\overline{G}^{(a+b)}(^1S_0)\Big{|}_{|\textbf{p}_{i,f}|} & = & c_f^{(2g)} \alpha_s^2
\Big [\frac{1}{m^2}c_0^{full}+\frac{1}{m^4}c_2^{full}+\frac{1}{m^6}c_4^{full}+\frac{1}{m^8}c_6^{full} + \textrm{higher order} \Big],
\end{eqnarray}
where $\overline{G}$ refers to the result of $G$ after the replacement, the subindexes $|\textbf{p}_{i,f}|$ mean to expand the expression on $|\textbf{p}_{i,f}|$ and $c^{full}_i$ are some functions on $p^2$ with corresponding orders of $|\textbf{p}_{i,f}|$. The manifest expressions of $c_{i}^{full}$ are a little complex and we list them in the Appendix. To compare the results with those given in NRQCD, we further expand the above results on $p^2$ and rearrange the results as
\begin{eqnarray}
\overline{G}^{(a+b)}(^1S_0) \Big{|}_{|\textbf{p}_i|,|\textbf{p}_f| ,p^2} & = & c_f^{(2g)}\alpha_s^2
\Big [ \frac{1}{m^2}c_0+\frac{1}{m^4}c_2+\frac{1}{m^6}c_4+\frac{1}{m^8}c_6 + \textrm{higher order} \Big ],
\end{eqnarray}
where $c_i$ are the combinations of terms with corresponding orders of $\textbf{p}_{i,f}^2$ and $p^2$. For $p^2>0$ $c_i$ are expressed as
\begin{eqnarray}
\textrm{Im}[c_0]&=& 2 \pi  ,\nonumber \\
\textrm{Im}[c_2] &=&-\frac{2\pi}{3}  \Big[5(|\textbf{p}_f|^2+|\textbf{p}_i|^2)-6 p^2 \Big],\nonumber \\
\textrm{Im}[c_4]&=&\frac{\pi }{360}  \Big[1059(|\textbf{p}_i|^4|+\textbf{p}_f|^4)+1640|\textbf{p}_i|^2|\textbf{p}_f|^2-1110(|\textbf{p}_i|^2|+|\textbf{p}_f|^2)p^2-450p^4\Big],\nonumber\\
\textrm{Im}[c_6] &=&-\frac{\pi}{10080} \Big[39435 (|\textbf{p}_i|^6+|\textbf{p}_f|^6)+42700 (|\textbf{p}_f|^4|\textbf{p}_f|^2|+\textbf{p}_f|^2|\textbf{p}_f|^4)
\nonumber \\&&~~~~~~~-55986(|\textbf{p}_f|^4+|\textbf{p}_f|^4|)p^2-21840|\textbf{p}_i|^2|\textbf{p}_f|^2p^2
\nonumber \\&&~~~~~~~+4935(|\textbf{p}_f|^2+|\textbf{p}_f|^2) p^4-7560 p^6)\Big] ,\nonumber \\
\end{eqnarray}

and
\begin{eqnarray}
\textrm{Re}[c_0]&=& 4  (1-\log 2) ,\nonumber \\
\textrm{Re}[c_2]&=&\frac{1}{3}  \Big[(20 \log 2-13)(|\textbf{p}_i|^2+|\textbf{p}_f|^2)+6(1- 4\log 2) p^2 \Big],\nonumber \\
\textrm{Re}[c_4]&=&\frac{1}{180} \Big[(180 \log r_{p^2}+1131-1419 \log 2)(|\textbf{p}_i|^4+|\textbf{p}_f|^4)
\nonumber\\&&~~~~~+\big[180 \log r_{p^2}+1010-2000 \log 2)|\textbf{p}_i|^2|\textbf{p}_f|^2
\nonumber\\&&~~~~~+(-540 \log r_{p^2}-510+2190 \log 2)(|\textbf{p}_i|^2+|\textbf{p}_f|^2)p^2
\nonumber\\&&~~~~~+(540 \log r_{p^2}-450-630\log 2)p^4
\nonumber\\&&~~~~~-\frac{45 (|\textbf{p}_f|^6+|\textbf{p}_f|^4 |\textbf{p}_i|^2+|\textbf{p}_f|^2 |\textbf{p}_i|^4+|\textbf{p}_i|^6)}{p^2} \Big], \nonumber\\
\textrm{Re}[c_6] &=&\frac{1}{5040}\Big[ (1680 \log r_{p^2}+25968-42795 \log 2) (|\textbf{p}_i|^6+|\textbf{p}_f|^6)
\nonumber\\&&~~~~~+(11760 \log r_{p^2}+40047-66220 \log 2)(|\textbf{p}_i|^4|\textbf{p}_f|^2+|\textbf{p}_i|^4|\textbf{p}_f|^2 )
\nonumber\\&&~~~~~+(-6720 \log r_{p^2}-5040+69426 \log2)(|\textbf{p}_i|^4+|\textbf{p}_f|^4)p^2
\nonumber\\&&~~~~~+(-35280 \log r_{p^2}-16380+92400 \log2)|\textbf{p}_i|^2|\textbf{p}_f|^2p^2
\nonumber\\&&~~~~~+(8400 \log r_{p^2}-21735-21735 \log2)(|\textbf{p}_i|^2+|\textbf{p}_f|^2)p^4
\nonumber\\&&~~~~~+(5040 \log r_{p^2}+2100-1680 \log2)p^6
\Big ],
\label{eq:ci-expand}
\end{eqnarray}
where $r_{p^2}\triangleq p^2/m^2$ and  is assumed to be larger than 0 in the above expressions. For the $r_{p^2}<0$ case, the term $\log r_{p^2}$ should be taken as $\log(r_{p^2}+i\epsilon)$ with $\epsilon=0^+$ and it gives an additional contribution to the imagine part of the coefficients. This analytic continuation is also checked by the direct calculation with $r_{p^2}<0$. For convenient we also list the expressions with $p^2<0$ which can be used in the positronium system.
\begin{eqnarray}
\textrm{Im}[c_0^{(p^2<0)}]&=& \textrm{Im}[c_0],\nonumber \\
\textrm{Im}[c_2^{(p^2<0)}] &=&\textrm{Im}[c_2],\nonumber \\
\textrm{Im}[c_4^{(p^2<0)}]&=&\frac{\pi }{360}  \Big[1419(|\textbf{p}_i|^4|+\textbf{p}_f|^4)+2000|\textbf{p}_i|^2|\textbf{p}_f|^2-2190(|\textbf{p}_i|^2|+|\textbf{p}_f|^2)p^2+630p^4\Big],\nonumber\\
\textrm{Im}[c_6^{(p^2<0)}] &=& -\frac{\pi}{10080} \Big[42795 (|\textbf{p}_i|^6+|\textbf{p}_f|^6)+66220 (|\textbf{p}_f|^4|\textbf{p}_f|^2|+\textbf{p}_f|^2|\textbf{p}_f|^4)
\nonumber \\&&~~~~~~~-69426(|\textbf{p}_f|^4+|\textbf{p}_f|^4|)p^2-92400|\textbf{p}_i|^2|\textbf{p}_f|^2p^2
\nonumber \\&&~~~~~~~+21735(|\textbf{p}_f|^2+|\textbf{p}_f|^2) p^4+2520 p^6)\Big].
\label{eq:ci-expand-below}
\end{eqnarray}
The real parts of the expressions for $p^2<0$ are same with the expressions with $p^2>0$ by changing $r_p^2$ to $-r_p^2$.

Using the above expressions and the quasi potential method, one can directly get the corresponding non-relativistic potential in the LO-$\alpha_s$. In the momentums space it is expressed as
\begin{eqnarray}
V_{eff}(^1S_0)&=& - G^{(a+b)}(^1S_0),
\end{eqnarray}
since we have normalized the Dirac spinors as $u^+u =v^+v= 1$.

Taking this effective potential as a perturbative interaction comparing with the non-perturbative potential in the quark model. The corresponding energy shift in the leading order can be got directly as
\begin{eqnarray}
\Delta E (s) &=& <V_{eff}^{^1S_0}> \nonumber\\
&=&  \int d^3\textbf{p}_id^3\textbf{p}_f\Phi^*(\textbf{p}_f) V_{eff}(^1S_0,p_i^2,p_f^2,p_i\cdot p_f,s)\Phi(\textbf{p}_i),
\end{eqnarray}
where $\Phi(\textbf{p}_{i,f})\triangleq \phi(|\textbf{p}_{i,f}|)Y_{00}$  are the wave functions of the $^1S_0$ states in the momentum space normalized as $\int d^3\textbf{p}\Phi^*(\textbf{p})\Phi(\textbf{p})=1$.


Corresponding, the result after the expansion and the integration is expressed as
\begin{eqnarray}
\Delta E (s)\big{|}_{|\textbf{p}_i|,|\textbf{p}_f|} &=& -2\pi^3c_f^{(2g)} \alpha_s^2 \Big [ \frac{c_{0 1}^{full}}{m^2}\psi(0)^2-\frac{c_{21}^{full}}{m^4}2\psi^{(2)}(0)\psi(0)+\frac{c_{41}^{full}}{m^6}2\psi^{(4)}(0)\psi(0)
\nonumber\\
&&~~+\frac{c_{42}^{full}}{m^6}\psi^{(2)}(0)^2-\frac{c_{61}^{full}}{m^8}2\psi^{(6)}(0)\psi(0)
-\frac{c_{62}^{full}}{m^8}2\psi^{(2)}(0)\psi^{(4)}(0) \Big],
\label{eq:energy-result}
\end{eqnarray}

where the coefficients can be found in the Appendix and $\psi^{(n)}(0)$ are defined as
\begin{eqnarray}
\psi^{(n)}(0)& \triangleq & \frac{(-i)^n}{(2\pi)^{3/2}} \int |\textbf{p}|^{n}\Phi(\textbf{p}) d^3\textbf{p},
\end{eqnarray}
with $n$ even. $\psi^{(n)}(0)$ are corresponding to the values of the $n$-th derivative of the wave functions in the coordinate space at $r=0$ since
\begin{eqnarray}
\Phi(\textbf{p})& = &  \int \frac{1}{(2\pi)^{3/2}}e^{-i \textbf{p} \cdot \textbf{r}} \psi(\textbf{r}) d^3\textbf{r}.
\end{eqnarray}
One should note that the angle part $Y_{00}$ is included in the wave function $\psi(\textbf{r})$.

Furthermore, one can expand the result on $p^2$ which gives
\begin{eqnarray}
\Delta E (s)\big{|}_{|\textbf{p}_i|,|\textbf{p}_f|,p^2} &=&  - 2\pi^3 c_f^{(2g)} \alpha_s^2\Big [\frac{c_0}{m^2}\psi(0)^2+ \textrm{higher terms}\Big].
\end{eqnarray}
Physically, the imagine part of $\Delta E(s)$ is corresponding to the decay width of a state as $\Gamma =-2\textrm{Im}[\Delta E(s)]$  with $\sqrt{s}= M_{^1S_0}$ and the real part of $\Delta E(s)$ is corresponding to the mass shift due to the two-gluon annihilation in the leading order.

Eq. (\ref{eq:energy-result}) can also be  directly used to estimate the decay widths of $\eta_{c,b}\rightarrow2\gamma$, the mass shifts and decay widths of positronium in $^1S_0$ states due to the annihilation effect $e^+e^-\rightarrow2\gamma$. For $\eta_{c,b}\rightarrow2\gamma$, we should replace the factor $c_f^{(2g)}\alpha_s^2$ by $c_f^{(2\gamma)}Q_{c,b}^4\alpha_{QED}^2$ with $c_f^{(2\gamma)}=3$ and $Q_{b,c}$ the electric charges of $b,c$ quarks. For positronium, we should replace the factor $c_f^{(2g)}\alpha_s^2$ by $\alpha_{QED}^2$. Since in the positronium case one has $r_{p^2}<0$ or $r_s\triangleq \sqrt{s}/2m <1$, the corresponding $\textrm{Im}[c_i^{full}]$ and $\textrm{Re}[c_i^{full}]$ should be re-written by changing $\log(r_s^2-1)$ to $\log(r_s^2-1+i\epsilon)$ like Eq. (\ref{eq:ci-expand-below}).

\section{\label{sec-3} comparison with the results of NRQCD in LO-$\alpha_s$}

To compare the above results with the corresponding results in NRQCD in LO-$\alpha_s$, we can apply the on shell constrain conditions to the coefficients $c_i$. The on shell constrain conditions means $|\textbf{p}_i|^2=|\textbf{p}_f|^2=-p_i^2=-p_f^2=p^2$ which leads to the following results.
\begin{eqnarray}
\textrm{Im}[c_0^{(on)}]&=&2 \pi  ,\nonumber \\
\textrm{Im}[c_2^{(on)}] &=&-\frac{8\pi  }{3}  |\textbf{p}_i|^2,\nonumber \\
\textrm{Im}[c_4^{(on)}]&=&\frac{136\pi  }{45}   |\textbf{p}_i|^4,\nonumber \\
\textrm{Im}[c_6^{(on)}] &=& -\frac{1024\pi }{315} |\textbf{p}_i|^6,
\label{eq:on-shell-Im}
\end{eqnarray}
where the index $(on)$ refers to the results after applying the  on shell conditions to $c_i$.

Comparing these results to the decay width of  $\Gamma(\eta_c \rightarrow 2\gamma)$ in NRQCD \cite{project-operator-2} which expressed as
\begin{eqnarray}
&&\Gamma(H(^1S_0) \rightarrow \gamma \gamma) \nonumber\\
=&&\frac{2\alpha_{QED}^2 Q^4 \pi}{m^2}\langle H(^1S_0) | \mathcal
O_{\textrm{em}} (^1S_0)| H(^1S_0)\rangle - \frac{8\alpha_{QED}^2 Q^4 \pi}{3 m^4}\langle H(^1S_0) | \mathcal
P_{\textrm{em}} (^1S_0)| H(^1S_0)\rangle \nonumber\\
&&+\frac{20\alpha_{QED}^2 Q^4 \pi}{9 m^6}\langle H(^1S_0) | \mathcal
Q'_{\textrm{em}} (^1S_0)| H(^1S_0)\rangle
+\frac{4\alpha_{QED}^2 Q^4  \pi }{5 m^6}\langle H(^1S_0) | \mathcal
Q''_{\textrm{em}} (^1S_0)| H(^1S_0)\rangle,\nonumber \\
\label{NQRCD}
\end{eqnarray}
with
\begin{eqnarray}
\mathcal{O}_{\textrm{em}}(^1S_0) &= &
\psi^{\dag} \chi |0 \rangle \langle 0|\chi^{\dag}\psi, \nonumber \\
\mathcal{P}_{\textrm{em}}(^1S_0) &= &
\frac{1}{2}
\psi^{\dag}(-\frac{i}{2}\overleftrightarrow D )^2\chi |0 \rangle
\langle 0|  \chi^{\dag}\psi+ \textrm{H.c.}, \nonumber \\
\mathcal{Q'}_{\textrm{em}}(^1S_0) &= &
\psi^{\dag}(-\frac{i}{2}\overleftrightarrow D )^2\chi|0
\rangle \langle 0|\chi^{\dag}(-\frac{i}{2}\overleftrightarrow D )^2\psi,\nonumber \\
\mathcal{Q''}_{\textrm{em}}(^1S_0) &= &
\frac{1}{2}
\psi^{\dag}(-\frac{i}{2}\overleftrightarrow D )^4\chi  |0 \rangle
\langle 0|\chi^{\dag}\psi
+ \textrm{H.c.}.
\label{NQRCD-operator}
\end{eqnarray}
One can find the $\textbf{p}_{i,f}$ in Eq. (\ref{eq:on-shell-Im}) are correcponding to the operaor $-\frac{i}{2}\overleftrightarrow D$ in Eq. (\ref{NQRCD-operator}), $\textrm{Im}[c^{(on)}_{0,2}]$ are same with the coefficients of the 1st, 2nd terms  and $\textrm{Im}[c^{(on)}_4]$ is same with the sum of the coefficients of 3rd and 4th terms of Eq.(\ref{NQRCD}) except a global factor $\alpha_{QED}^2Q^4$ and a normalized factor $1/m^n$.  This is natural since we just take $s$ as a free variable at first and then apply the on shell conditions after the loop integration, the results should go back to  NRQCD results which takes the on shell conditions directly. The coefficient $\textrm{Im}[c^{(on)}_6]$ is corresponding to the sum of coefficients in NRQCD in order $v^6$ \cite{JiaYu2011}. In our calculation we take the quantities $p^2, |\textbf{p}_i|,|\textbf{p}_f|,\textbf{p}_i\cdot\textbf{p}_f$ as independent variables at first and then using the on shell conditions $p^2=|\textbf{p}_i|^2=|\textbf{p}_f|^2$, this means that we can not distinguish $|\textbf{p}_i|$ and $|\textbf{p}_f|$ when applying the on shell conditions and we can only give the sum of the corresponding NRQCD coefficients. The real part of $c_i^{(on)}$ can also be got easily from Eq.(\ref{eq:ci-expand}) as following.
\begin{eqnarray}
\textrm{Re}[c_0^{(on)}]&=&4 (1-\log2) ,\nonumber \\
\textrm{Re}[c_2^{(on)}] &=&\frac{4}{3}  |\textbf{p}_i|^2 (4\log2-5) ,\nonumber \\
\textrm{Re}[c_4^{(on)}]&=&\frac{1}{90} |\textbf{p}_i|^4 (811-544 \log2) ,\nonumber \\
\textrm{Re}[c_6^{(on)}] &=& \frac{1}{630}  |\textbf{p}_i|^6(4096\log2-8025),
\label{eq:on-shell-Re}
\end{eqnarray}
where the result $\textrm{Re}[c_0^{(on)}]$ is same with that given in the reference \cite{Real-part-refs}. We want to point out that for positronium where $p^2<0, r_s<1$, we do not suggest use the Eq. (\ref{eq:on-shell-Im},\ref{eq:on-shell-Re}) but suggest to use the corresponding expressions with $p^2<0$.

\section{The numerical result and conclusion}
\begin{figure*}[htbp]
\center{\epsfxsize 3.5 truein\epsfbox{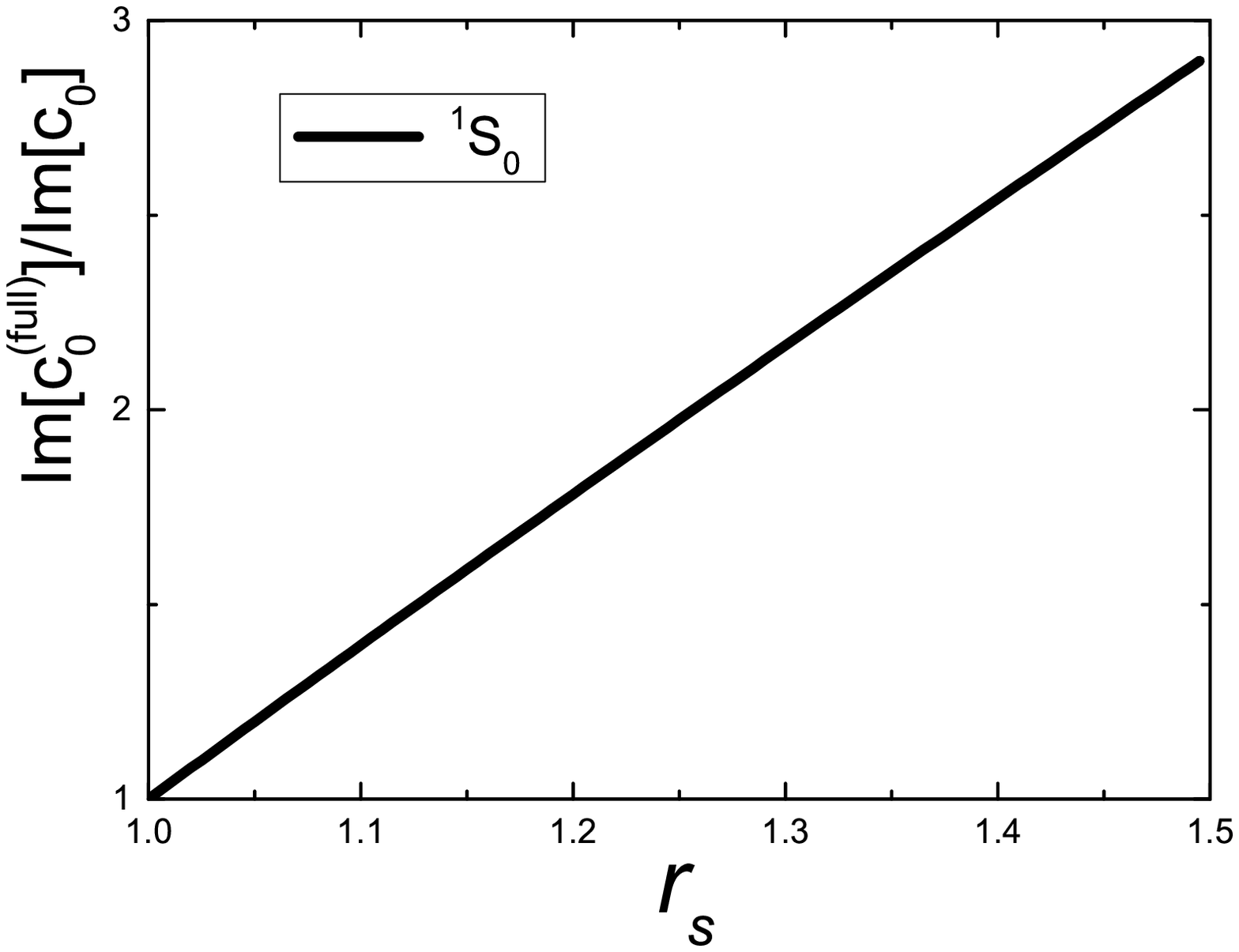}\epsfxsize 3.5 truein\epsfbox{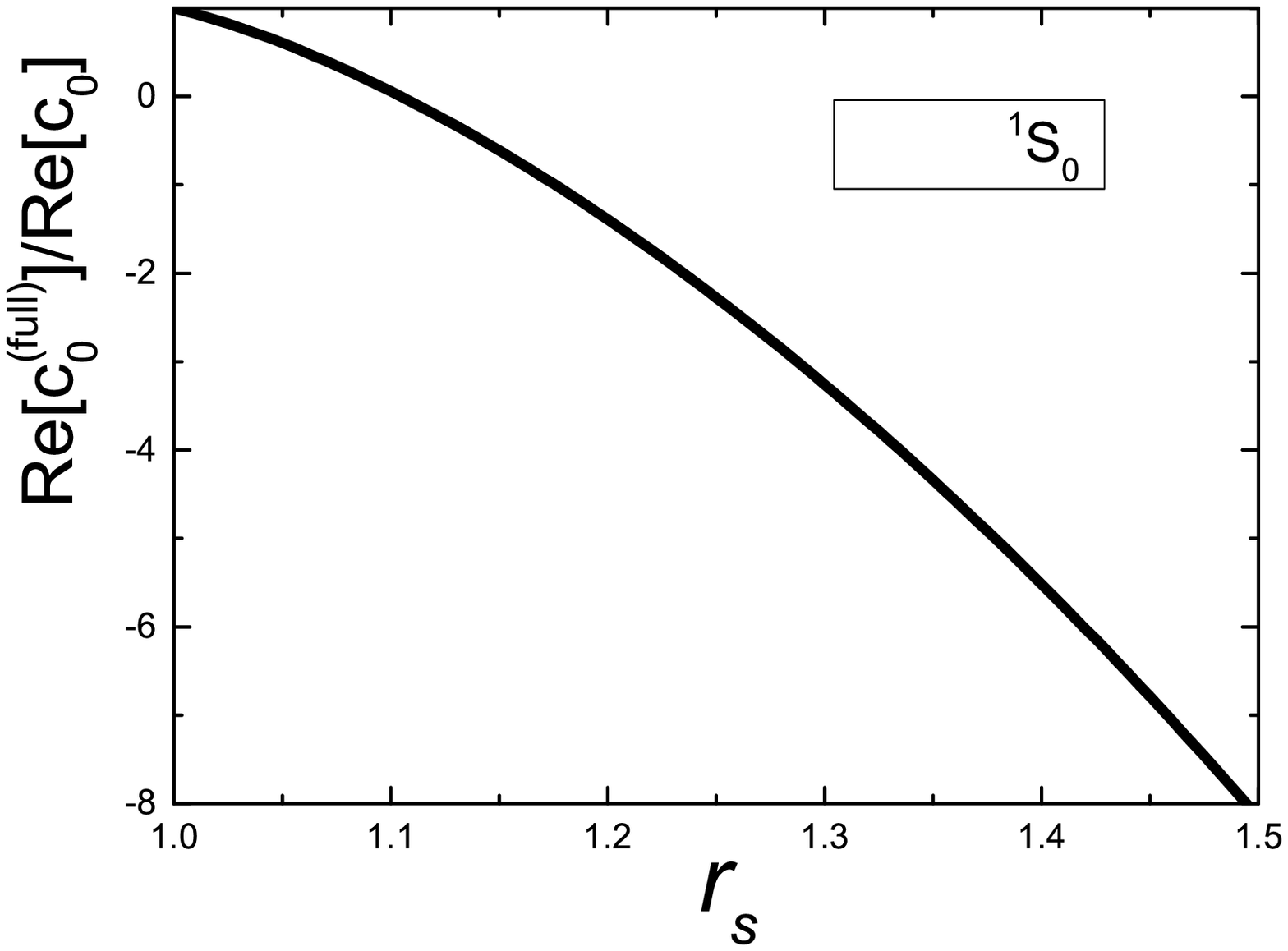}}
\caption{The numerical results for the corrections to the  decay width and the mass shift for $^1S_0$ state in the leading order of $\alpha_s$ and $v$ where $r_s \triangleq \sqrt{s}/2m$.}
\label{figure:ratio-off-on-1S0}
\end{figure*}

For the visualization, we list some numerical results in this section. In NRQCD, the contribution to the decay width in the LO-$v$ and LO-$\alpha_s$ is determined by the coefficient $c_0$ and the wave function at zero point $\psi(0)$. In our calculation, we expand the expression only on $|\textbf{p}_i|, |\textbf{p}_f|$ and $\textbf{p}_i \cdot \textbf{p}_f$ and do not take the on shell conditions to fix $s$. This results in the corresponding decay width is also dependent on $s$. The ratio $\textrm{Im}[c_0^{full}]/\textrm{Im}[c_0]$ reflects the corresponding correction to the decay width in NRQCD in leading order due to the off shell effect. Such correction is not dependent on the non-perturbative parameter $\psi(0)$ but only dependent on the ratio $r_s$. The corresponding numerical results are presented in the left panel of Fig. \ref{figure:ratio-off-on-1S0}.

In the right panel of Fig. \ref{figure:ratio-off-on-1S0}, we also present the similar correction to the real part of the coefficient. The ratio $\textrm{Re}[c_0^{full}]/\textrm{Re}[c_0]$ reflects the correction to the mass shift due to the off shell effect.  In the  positronium case,  the bound energies of the physical bound states are very small relative to the mass of electron which means the ratios $|r_s|=(2m-\Delta E)/2m$ are very close to 1 and the corrections are very small. In the charmonium case, the bound energies of the physical states are not so small comparing with the quark mass and the $r_s$ for the real physical states are also not small. For
example, if we take the mass of $c$ quark as a constant, then the relative ratio $r_s(\eta_{c}(2S))/r_s(\eta_{c}(1S))=M_{\eta_c(2S)}/M_{\eta_c(1S)}=1.22$ which is not a small value. This means the correction is not small.
\begin{figure*}[htbp]
\center{\epsfxsize 3.5 truein\epsfbox{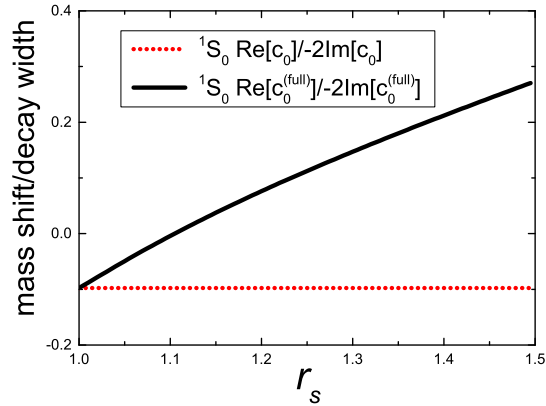}}
\caption{The numerical results for the ratio of the mass shift and the decay width for $^1S_0$ state in the leading order of $\alpha_s$ and $v$ .}
\label{figure:ratio-Re-Im-1S0}
\end{figure*}

In Fig. \ref{figure:ratio-Re-Im-1S0}, the ratios $\textrm{Re}[c_0^{full}]/-2\textrm{Im}[c_0^{full}] $ and $\textrm{Re}[c^{(on)}_0]/-2\textrm{Im}[c^{(on)}_0]$ are presented. These ratios reflect the relation between the mass shifts and the decay widths. If one assumes the LO gives the most of the contribution, one can use the experimental decay width to estimate the mass shift. For the heavy quarknia, if we take the approximation $r_s(\eta_c)\approx 1$ and $r_s(\eta_b)\approx 1$ then the corrections to the mass shifts are about $-(1-\log2)/\pi \Gamma(\eta_c) \approx -3.1$ MeV for $\eta_c$, $0.09 \Gamma(\eta_c(2S)) = 1.0$ MeV for $ \eta_c(2S)$, $-(1-\log2)/\pi \Gamma(\eta_b) \approx -6.1$ MeV for $\eta_b$ and $-0.04 \Gamma(\eta_b(2S)) $ for $ \eta_b(2S)$. The interesting property is that the mass shifts are strong dependent on the mass or the binding energy of $^1S_0$ states. This means the corrections to the different $^1S_0$ states are very different and can not be subtracted or hidden in a unified way.

In summary, the non-relativistic asymptotic behavior of the transition $q\overline{q}\rightarrow2g\rightarrow q\overline{q}$ in the $^1S_0$ channel is discussed. In our discussion, the momenta of the quarks and anti-quarks are not limited on mass shell after projecting the quark anti-quark pairs to $^1S_0$ state. We calculate the results by expanding the expression on the three-dimensional momenta of quarks and anti-quarks to order 6. The imagine part of the first 3 terms of our results after applying the on shell conditions can reproduce the non-relativistic QCD (NRQCD) results in leading order of $\alpha_s$. The real part of our results can be used to estimate the mass shift of  $^1S_0$ heavy quark anti-quark system due to the $2g$ annihilation effect. The results can also be used to estimate the energy shifts of $^1S_0$ states of positronium.

\section{Acknowledgments}
The author Hai-Qing Zhou would like to thank Wen-Long Sang, Zhi-Yong Zhou and Dian-Yong Chen for their kind and helpful discussions. This work is supported by the  National Natural Science Foundations of China under Grant No. 11375044.

\section{Appendix}
In this Appendix, the expressions for $c_i^{full}$ are listed.

\begin{eqnarray}
c_0^{full}&=& c_{01}^{full},\nonumber
\end{eqnarray}
\begin{eqnarray}
c_2^{full}&=& (|\textbf{p}_i|^2+|\textbf{p}_f|^2) c_{21}^{full},\nonumber
\end{eqnarray}
\begin{eqnarray}
c_4^{full}&=& (|\textbf{p}_i|^4+|\textbf{p}_f|^4)c_{41}^{full}+ |\textbf{p}_i|^2|\textbf{p}_f|^2 c_{42}^{full},\nonumber
\end{eqnarray}
\begin{eqnarray}
c_6^{full}&=& (|\textbf{p}_i|^6+|\textbf{p}_f|^6)c_{61}^{full}+ (|\textbf{p}_i|^4|\textbf{p}_f|^2+ |\textbf{p}_i|^2|\textbf{p}_f|^4)c_{62}^{full},
\end{eqnarray}
with

\begin{eqnarray}
\textrm{Im}[c_{01}^{full}]&=&\frac{\pi  \alpha ^2 (\text{rs}+1)^4(3 r_s^4-1)}{4 \text{mc}^2
   r_s^2 (r_s^2+1)^2},
\end{eqnarray}

\begin{eqnarray}
\textrm{Im}[c_{21}^{full}]&=&-\frac{\pi(r_s+1)^2}{48m^4 r_s^4 (r_s^2+1)^4} \Big(35 r_s^{12}+94 r_s^{11}+145r_s^{10}+100 r_s^9+172 r_s^8
\nonumber\\&&~~+52 r_s^7+36 r_s^6-12r_s^5-31 r_s^4+22 r_s^3+3 r_s^2+16 r_s+8\Big),
\end{eqnarray}

\begin{eqnarray}
\textrm{Im}[c_{41}^{full}]&=&\frac{\pi  (r_s+1)^2}{240 m^6 r_s^6 (r_s^2+1)^6}\Big(185 r_s^{18}+520 r_s^{17}+1139 r_s^{16}+1518 r_s^{15}+2566 r_s^{14}
\nonumber\\&&~~+2404 r_s^{13}+2756 r_s^{12}+888 r_s^{11}+1329 r_s^{10}+20 r_s^9+15 r_s^8-510 r_s^7
\nonumber\\&&~~-284 r_s^6-504 r_s^5-210 r_s^4-296 r_s^3-132 r_s^2-72 r_s-36\Big),
\end{eqnarray}

\begin{eqnarray}
\textrm{Im}[c_{42}^{full}]&=&\frac{\pi}{576 m^6 r_s^6 (r_s^2+1)^6}  \Big(375 r_s^{20}+2220 r_s^{19}+6102 r_s^{18}+11084 r_s^{17}+16964 r_s^{16}
\nonumber\\&&~~+20820 r_s^{15}+24114 r_s^{14}+19988 r_s^{13}+18494 r_s^{12}+11892 r_s^{11}+10810 r_s^{10}
\nonumber\\&&~~+7060 r_s^9+5484 r_s^8+5772 r_s^7+3230 r_s^6+2892 r_s^5+843 r_s^4+128 r_s^3
\nonumber\\&&~~-96 r_s^2-192 r_s-48\Big),
\end{eqnarray}

\begin{eqnarray}
\textrm{Im}[c_{61}^{full}]&=&-\frac{\pi   (r_s+1)^2}{26880 m^8 r_s^8 (r_s^2+1)^8} \Big(21595 r_s^{24}+61670 r_s^{23}+176925 r_s^{22}+300580 r_s^{21}
\nonumber\\&&~~+590734 r_s^{20}+710200 r_s^{19}+1068266 r_s^{18}+768220 r_s^{17}+1099948 r_s^{16}
\nonumber\\&&~~+608700 r_s^{15}+596964 r_s^{14}+50540 r_s^{13}+95578 r_s^{12}+1960 r_s^{11}-54698 r_s^{10}
\nonumber\\&&~~+106820 r_s^9+10409 r_s^8+177310 r_s^7+69135 r_s^6+128720 r_s^5+57576 r_s^4
\nonumber\\&&~~+48480 r_s^3+23088 r_s^2+7680 r_s+3840\Big),
\end{eqnarray}

\begin{eqnarray}
\textrm{Im}[c_{62}^{full}] &=& -\frac{\pi}{2880 m^8 r_s^8 \left(r_s^2+1\right)^8}  \Big(1950 r_s^{26}+12030 r_s^{25}+36570 r_s^{24}+82940 r_s^{23}
\nonumber\\&&~~+154123 r_s^{22}+244280 r_s^{21}+346088 r_s^{20}+404804 r_s^{19}+451148 r_s^{18}
\nonumber\\&&~~+403236 r_s^{17}+389788 r_s^{16}+252548 r_s^{15}+208854 r_s^{14}+107264 r_s^{13}
\nonumber\\&&~~+76392 r_s^{12}+18460 r_s^{11}-1386 r_s^{10}-22594 r_s^9-16038 r_s^8-20096 r_s^7
\nonumber\\&&~~-6889 r_s^6-4504 r_s^5+416 r_s^4+1824 r_s^3+1032 r_s^2+768 r_s+192\Big)
  \end{eqnarray}

\begin{eqnarray}
\textrm{Re}[c_{01}^{full}]&=& -\frac{ (r_s+1)^4 }{4 m^2 r_s^2 (r_s^2+1)^2} \Big [4 r_s^6 \log r_s+r_s^6 (4\log2-1)-2 r_s^4-r_s^2
\nonumber\\&&~~+(-2 r_s^6+3 r_s^4-1 )\log (r_s^2-1)\Big],
\end{eqnarray}

\begin{eqnarray}
\textrm{Re}[c_{21}^{full}] &=& \frac{(r_s+1)^2 }{48 m^4 r_s^4 (r_s^2+1)^4} \Big \{4 ( 15 r_s^6+30 r_s^5+61 r_s^4+52 r_s^3+77 r_s^2+54 r_s+31 ) r_s^8 \log r_s
\nonumber\\&&~~-(r_s+1)^2 (30 r_s^{12}+57 r_s^{10}-104 r_s^9+160 r_s^8-208 r_s^7+146 r_s^6-136 r_s^5
\nonumber\\&&~~+90 r_s^4-32 r_s^3+5 r_s^2-8 ) \log (r_s^2-1)+r_s^2 \Big [15 r_s^{12} (4\log2-1)
\nonumber\\&&~~+30 r_s^{11} (4\log2-1)+r_s^{10} (244 \log2-85)+4 r_s^9 (52 \log2-31)
\nonumber\\&&~~+14 r_s^8 (22 \log2-13)+8 r_s^7 (27 \log2-20)+2 r_s^6 (62 \log2-89)
\nonumber\\&&~~-52 r_s^5-67 r_s^4+30 r_s^3+7 r_s^2+16 r_s+8 \Big ]\Big \},
\end{eqnarray}

\begin{eqnarray}
\textrm{Re}[c_{41}^{full}] &=&-\frac{ (r_s+1)^2}{480 m^6 r_s^6 (r_s^2+1)^6} \Big \{4 (165 r_s^{10}+330 r_s^9+1025 r_s^8+1220 r_s^7+2482 r_s^6
\nonumber\\&&~~+2128 r_s^5+3074 r_s^4+1340 r_s^3+1977 r_s^2+870 r_s+525) r_s^{10} \log r_s
\nonumber\\&&~~+r_s^2 \Big [165 r_s^{18} (4\log2-1)+330 r_s^{17} (4\log2-1)
\nonumber\\&&~~+2 r_s^{16} (2050 \log2-647)+r_s^{15} (4880 \log2-2018)
\nonumber\\&&~~+r_s^{14} (9928 \log2-4215)+28 r_s^{13} (304 \log2-165)
\nonumber\\&&~~+4 r_s^{12} (3074 \log2-1829)+20 r_s^{11} (268 \log2-283)
\nonumber\\&&~~+r_s^{10} (7908 \log2-7231)+6 r_s^9 (580 \log2-723)+6 r_s^8 (350 \log2-681)
\nonumber\\&&~~-2442 r_s^7-1433 r_s^6-1352 r_s^5-576 r_s^4-664 r_s^3-300 r_s^2-144 r_s-72 \Big ]
\nonumber\\&&~~-2 (165 r_s^{20}+330 r_s^{19}+840 r_s^{18}+700 r_s^{17}+1343 r_s^{16}+610 r_s^{15}+508 r_s^{14}
\nonumber\\&&~~-1064 r_s^{13}-779 r_s^{12}-18 r_s^{11}-804 r_s^{10}-20 r_s^9-15 r_s^8+510 r_s^7+284 r_s^6
\nonumber\\&&~~+504 r_s^5+210 r_s^4+296 r_s^3+132 r_s^2+72 r_s+36 ) \log (r_s^2-1)\Big \}
\end{eqnarray}

\begin{eqnarray}
\textrm{Re}[c_{42}^{full}] &=&-\frac{1}{576 m^6 r_s^6 (r_s^2+1)^6} \Big \{4 (15 r_s^6+30 r_s^5+61 r_s^4+52 r_s^3+77 r_s^2+54 r_s+31)^2 r_s^{10} \log r_s
\nonumber\\&&~~+r_s^2 \Big[225 r_s^{20} (4\log2-1)+900 r_s^{19} (4\log2-1)+8 r_s^{18} (1365 \log2-382)
\nonumber\\&&~~+4 r_s^{17} (5220 \log2-1771)+4 r_s^{16} (9151 \log2-3580)+4 r_s^{15} (12584 \log2-5645)
\nonumber\\&&~~+16 r_s^{14} (4067 \log2-1987)+4 r_s^{13} (16456 \log2-9159)+r_s^{12} (61308 \log2-37202)
\nonumber\\&&~~+4 r_s^{11} (11540 \log2-7531)+8 r_s^{10} (3845 \log2-2697)+36 r_s^9 (372 \log2-233)
\nonumber\\&&~~+4 r_s^8 (961 \log2-750)+4164 r_s^7+2528 r_s^6+2892 r_s^5+779 r_s^4+32 r_s^3
\nonumber\\&&~~-120 r_s^2-192 r_s-48\Big]-(r_s+1)^2 (450 r_s^{20}+900 r_s^{19}+2835 r_s^{18}
\nonumber\\&&~~+1650 r_s^{17}+6065 r_s^{16}+304 r_s^{15}+8899 r_s^{14}-6010 r_s^{13}+9661 r_s^{12}
\nonumber\\&&~~-10220 r_s^{11}+7665 r_s^{10}-10306 r_s^9+4059 r_s^8-4872 r_s^7+201 r_s^6
\nonumber\\&&~~-1302 r_s^5-827 r_s^4+64 r_s^3-144 r_s^2+96 r_s+48) \log (r_s^2-1)\Big\}
\end{eqnarray}

\begin{eqnarray}
\textrm{Re}[c_{61}^{full}]&=& \frac{(r_s+1)^2}{26880 m^8 (r_s-1) r_s^8 \left(r_s^2+1\right)^8}\Big \{4 (9765 r_s^{15}+9765 r_s^{14}+61145 r_s^{13}
\nonumber\\&&+30345 r_s^{12}+170205 r_s^{11}+8925 r_s^{10}+256329 r_s^9-147719 r_s^8
\nonumber\\&&+243559 r_s^7-269849 r_s^6+95091 r_s^5-349821 r_s^4+85911 r_s^3
\nonumber\\&&-135401 r_s^2-26005 r_s-42245) r_s^{12} \log r_s+r_s^2 \Big[9765 r_s^{25} (4\log2-1)
\nonumber\\&&+9765 r_s^{24}(4\log2-1)+r_s^{23} (244580 \log 2-77107)
\nonumber\\&&+r_s^{22} (121380 \log 2-61763)+20 r_s^{21} (34041 \log 2-12923)
\nonumber\\&&+100 r_s^{20} (357 \log 2-1123)+36 r_s^{19} (28481 \log 2-13805)
\nonumber\\&&+r_s^{18} (21980-590876 \log 2)+r_s^{17} (974236 \log 2-589950)
\nonumber\\&&+r_s^{16} (455250-1079396 \log 2)+r_s^{15} (380364 \log 2-493106)
\nonumber\\&&+r_s^{14} (856606-1399284 \log 2)+12 r_s^{13} (28637 \log 2-28349)
\nonumber\\&&+r_s^{12} (818068-541604 \log 2)-4 r_s^{11} (53649+26005 \log 2)
\nonumber\\&&+r_s^{10} (506316-168980 \log 2)-174949 r_s^9+225419 r_s^8-153171 r_s^7
\nonumber\\&&+48941 r_s^6-85120 r_s^5-18080 r_s^4-27312 r_s^3-17328 r_s^2-3840 r_s-3840\Big]
\nonumber\\&&-(19530 r_s^{27}+19530 r_s^{26}+100695 r_s^{25}+20615 r_s^{24}+225155 r_s^{23}-105805 r_s^{22}
\nonumber\\&&+222504 r_s^{21}-414904 r_s^{20}+129052 r_s^{19}-239652 r_s^{18}-141546 r_s^{17}-208394 r_s^{16}
\nonumber\\&&+183558 r_s^{15}+275622 r_s^{14}-97048 r_s^{13}+9128 r_s^{12}+56658 r_s^{11}-161518 r_s^{10}
\nonumber\\&&+96411 r_s^9-166901 r_s^8+108175 r_s^7-59585 r_s^6+71144 r_s^5+9096 r_s^4+25392 r_s^3
\nonumber\\&&+15408 r_s^2+3840 r_s+3840) \log (r_s^2-1)\Big \}
\end{eqnarray}

\begin{eqnarray}
\textrm{Re}[c_{62}^{full}]&=&\frac{\alpha ^2}{5760 m^8 (r_s-1) r_s^8 (r_s^2+1)^8} \Big\{4 (2475 r_s^{17}+7425 r_s^{16}+25440 r_s^{15}
\nonumber\\&&+42420 r_s^{14}+88460 r_s^{13}+102200 r_s^{12}+158232 r_s^{11}+104060 r_s^{10}
\nonumber\\&&+126082 r_s^9-7142 r_s^8+3152 r_s^7-146036 r_s^6-109620 r_s^5
\nonumber\\&&-154560 r_s^4-93896 r_s^3-93372 r_s^2-39045 r_s-16275) r_s^{12} \log r_s
\nonumber\\&&+r_s^2 \Big[2475 r_s^{27} (4\log2-1)+7425 r_s^{26} (4\log2-1)
\nonumber\\&&+r_s^{25} (101760 \log 2-29293)+r_s^{24} (169680 \log 2-59503)
\nonumber\\&&+2 r_s^{23} (176920 \log 2-67389)+2 r_s^{22} (204400 \log 2-93319)
\nonumber\\&&+8 r_s^{21} (79116 \log 2-37589)+8 r_s^{20} (52030 \log 2-35251)
\nonumber\\&&+r_s^{19} (504328 \log 2-335142)-2 r_s^{18} (73537+14284 \log 2)
\nonumber\\&&+2 r_s^{17} (6304 \log 2-54273)+r_s^{16} (164954-584144 \log 2)
\nonumber\\&&+r_s^{15} (160444-438480 \log 2)-644 r_s^{14} (960 \log 2-521)
\nonumber\\&&-8 r_s^{13} (46948 \log 2-26001)-8 r_s^{12} (46686 \log 2-30185)
\nonumber\\&&-3 r_s^{11} (52060 \log 2-38211)+r_s^{10} (77227-65100 \log 2)
\nonumber\\&&+54071 r_s^9+9837 r_s^8+30518 r_s^7+5906 r_s^6+8664 r_s^5+2424 r_s^4
\nonumber\\&&-2160 r_s^3-720 r_s^2-1152 r_s-384\Big]-2 \big(2475 r_s^{29}+7425 r_s^{28}
\nonumber\\&&+23490 r_s^{27}+32340 r_s^{26}+63920 r_s^{25}+55830 r_s^{24}+87049 r_s^{23}
\nonumber\\&&+13903 r_s^{22}+24274 r_s^{21}-65858 r_s^{20}-43192 r_s^{19}-98124 r_s^{18}
\nonumber\\&&-96172 r_s^{17}-17320 r_s^{16}-50202 r_s^{15}+8218 r_s^{14}-8173 r_s^{13}+41657 r_s^{12}
\nonumber\\&&+19846 r_s^{11}+21208 r_s^{10}-6556 r_s^9+4058 r_s^8-13207 r_s^7-2385 r_s^6
\nonumber\\&&-4920 r_s^5-1408 r_s^4+792 r_s^3+264 r_s^2+576 r_s+192\big) \log (r_s^2-1)\Big\},
\end{eqnarray}
where $r_s \triangleq \sqrt{s}/2m >1$ is assumed, for $r_s<1$ we can change $\log(r_s^2-1)$ by $\log(r_s^2-1+i\epsilon)$ and the real parts give some additional contributions to the imagine parts.

\end{document}